\documentclass[aps,prb,reprint,superscriptaddress]{revtex4-1}
\usepackage{graphicx}
\usepackage{dcolumn}
\usepackage{bm} 
\usepackage{color}
\usepackage{braket}
\usepackage{epstopdf}

\begin{document}


\title{Disorder from order among anisotropic next-nearest-neighbor Ising spin chains in SrHo$_2$O$_4$}

\author{J.-J. Wen}
\altaffiliation[Present address: ]{Department of Applied Physics, Stanford University, Stanford, CA 94305}
\affiliation{Institute for Quantum Matter and Department of Physics and Astronomy, The Johns Hopkins University, Baltimore, Maryland 21218, USA}

\author{W. Tian}
\affiliation{Quantum Condensed Matter Division, Oak Ridge National Laboratory, Oak Ridge, Tennessee 37831, USA}

\author{V. O. Garlea}
\affiliation{Quantum Condensed Matter Division, Oak Ridge National Laboratory, Oak Ridge, Tennessee 37831, USA}

\author{S. M. Koohpayeh}
\affiliation{Institute for Quantum Matter and Department of Physics and Astronomy, The Johns Hopkins University, Baltimore, Maryland 21218, USA}

\author{T. M. McQueen}
\affiliation{Institute for Quantum Matter and Department of Physics and Astronomy, The Johns Hopkins University, Baltimore, Maryland 21218, USA}
\affiliation{Department of Chemistry and Department of Materials Science and Engineering, The Johns Hopkins University, Baltimore, Maryland 21218, USA}

\author{H.-F. Li}
\affiliation{J$\ddot{u}$lich Centre for Neutron Science JCNS, Forschungszentrum J$\ddot{u}$lich GmbH, Outstation at Institut Laue-Langevin, Bo$\hat{\imath}$te Postale 156, F-38042 Grenoble Cedex 9, France}
\affiliation{Institut f$\ddot{u}$r Kristallographie der RWTH Aachen University, D-52056 Aachen, Germany}

\author{J.-Q. Yan}
\affiliation{Department of Materials Science and Engineering, The University of Tennessee, Knoxville, Tennessee 37996, USA}

\author{J. A. Rodriguez-Rivera}
\affiliation{NIST Center for Neutron Research, National Institute of Standards and Technology, Gaithersburg, Maryland 20899, USA}
\affiliation{Department of Materials Science and Engineering, University of Maryland, College Park, Maryland 20742, USA}

\author{D. Vaknin}
\affiliation{Ames Laboratory and Department of Physics and Astronomy, Iowa State University, Ames, Iowa 50011, USA}

\author{C. L. Broholm}
\affiliation{Institute for Quantum Matter and Department of Physics and Astronomy, The Johns Hopkins University, Baltimore, Maryland 21218, USA}
\affiliation{Quantum Condensed Matter Division, Oak Ridge National Laboratory, Oak Ridge, Tennessee 37831, USA}
\affiliation{NIST Center for Neutron Research, National Institute of Standards and Technology, Gaithersburg, Maryland 20899, USA}

\date{\today}

\begin{abstract}
     We describe why Ising spin chains with competing interactions in $\rm SrHo_2O_4$ segregate into ordered and disordered ensembles at low temperatures ($T$). Using elastic neutron scattering, magnetization, and specific heat measurements, the two distinct spin chains are inferred to have N\'eel  ($\uparrow\downarrow\uparrow\downarrow$) and double-N\'eel ($\uparrow\uparrow\downarrow\downarrow$) ground states respectively. Below $T_\mathrm{N}=0.68(2)$~K, the N\'eel chains develop three dimensional (3D) long range order (LRO), which arrests further thermal equilibration of the double-N\'eel chains so they remain in a disordered incommensurate state for $T$ below $T_\mathrm{S}= 0.52(2)$~K. $\rm SrHo_2O_4$ distills an important feature of incommensurate low dimensional magnetism: kinetically trapped topological defects in a quasi$-d-$dimensional spin system can preclude order in $d+1$ dimensions.
\end{abstract}


\maketitle

\section{Introduction}

An interesting byproduct of the intense pursuit of materials that can host spin-liquids has been the discovery of nominally pure crystalline solids with frozen short range correlated magnetism.\cite{Lee1996,Zaliznyak1999,Cava2001,Nakatsuji2005,Hiroi2009} In some cases quenched disorder simply alters the ground state and defines a short spin correlation length, but for materials such as two dimensional SCGO\cite{Lida2014} and $\rm NiGa_2S_4$\cite{Nakatsuji2005} where the spin correlation length is much shorter than a plausible impurity spacing such explanations seem untenable. Instead in the present study, we propose that spin disorder in frustrated low dimensional magnets can result from a complex thermalization process in the absence of quenched disorder.

Comprising two types of Ising spin chains with nearest neighbor ($J_1$) and next nearest neighbor ($J_2$) interactions [anisotropic next-nearest-neighbor (ANNNI) models\cite{[][{, and references therein.}]Selke1988213}] organized on a honeycomb-like lattice, $\rm SrHo_2O_4$ provides a striking example.\cite{Karunadasa2005,Ghosh2011,Hayes2012,Young2012,Young2013,Poole2014} We show the chains straddle the $J_2/J_1=1/2$ critical point so that ``red" chains have a ground state that doubles the unit cell ($\uparrow\downarrow\uparrow\downarrow$) while the ground state for ``blue" chains is a double-N\'eel state ($\uparrow\uparrow\downarrow\downarrow$) [as illustrated in Fig.~\ref{fig:1}(b)]. While  red chains develop 3D LRO,  blue chains in the very same crystal cease further equilibration towards their more complex ground state when the red spins saturate in an ordered state.

$\rm SrHo_2O_4$ belongs to a family of iso-structural rare-earth strontium oxides, $\rm SrRE_2O_4$.\cite{Karunadasa2005,Ghosh2011,Hayes2012,Young2012,Young2013,Poole2014,PhysRevB.78.184410,PhysRevB.84.174435,PhysRevB.86.064203,Petrenko2014,Li2014} Recent experiments on this series of materials have revealed low temperature magnetic states ranging from a disordered state in $\rm SrDy_2O_4$\cite{Poole2014}
to noncollinear 3D LRO in $\rm SrYb_2O_4$\cite{PhysRevB.86.064203}. A very unusual coexistence of a 3D LRO and 1D short range order (SRO) was discovered in polycrystalline and single crystalline samples of $\rm SrHo_2O_4$\cite{Young2012,Young2013,Poole2014} and $\rm SrEr_2O_4$\cite{PhysRevB.78.184410,PhysRevB.84.174435}. No explanation for the coexistence of two drastically different types of correlations over different length scales in the same crystal has so far been provided. With the additional experimental results and analysis presented here, we are able to provide an explanation for a partially ordered magnetic
state where quenched disorder does not play an essential role.

\section{Experimental Methods}

\subsection{Crystal structure and synthesis}

\begin{figure}[t]
    \includegraphics[width=3.3 in]{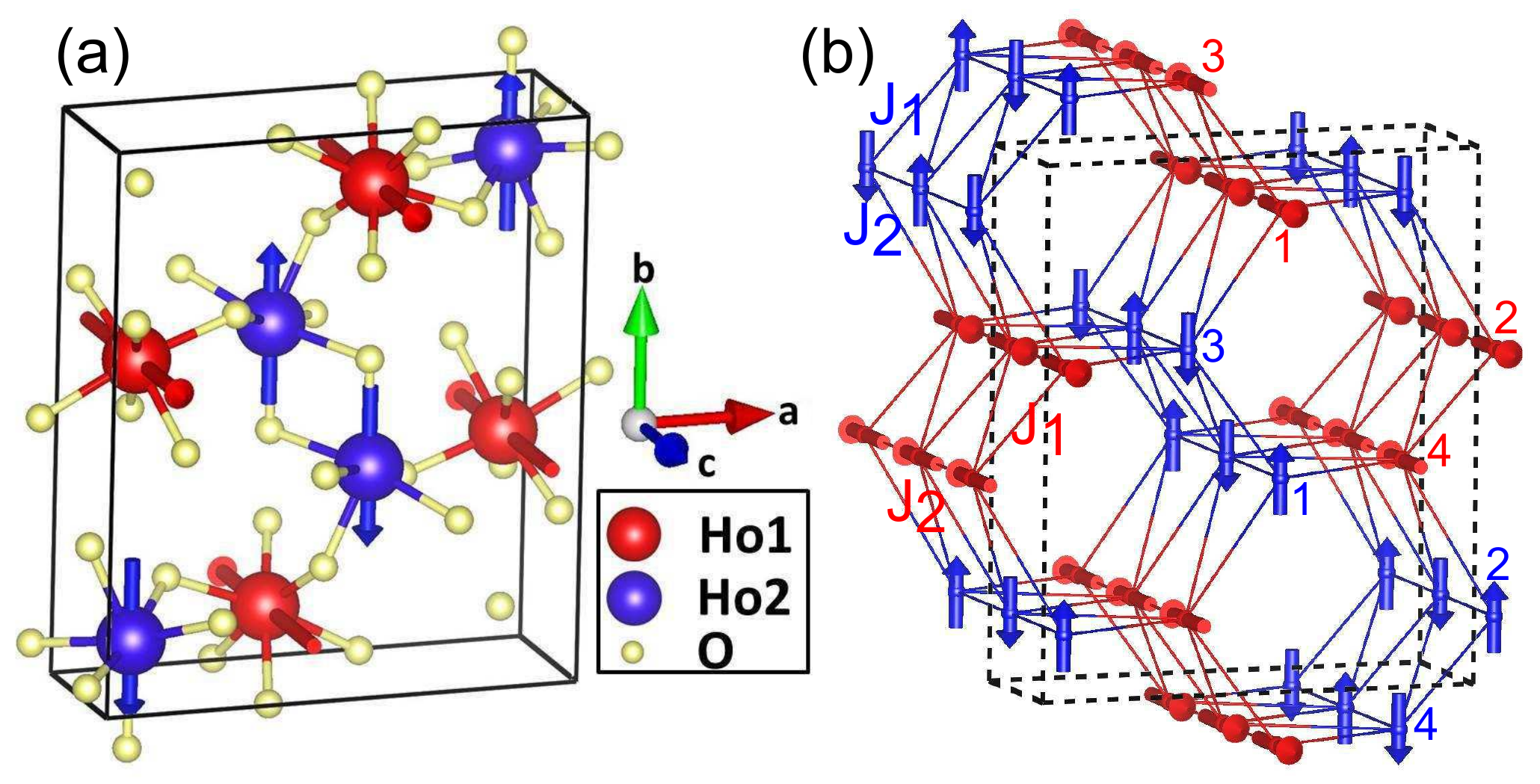}%
    \caption{\label{fig:1}(a) Crystallographic unit cell of SrHo$_2$O$_4$. Sr atoms were omitted for clarity. Red and Blue spheres show two distinct Ho sites, and the corresponding arrows show the Ising spin directions. (b) Magnetic lattice formed by Ho and a schematic representation of the spin structure determined by neutron diffraction.}
 \end{figure}

SrHo$_2$O$_4$ crystallizes in space group $Pnam$\cite{Karunadasa2005} with two inequivalent Ho sites [Fig.~\ref{fig:1}(a)]. Both are Wyckoff $4c$ sites with mirror planes perpendicular to the $\bf{c}$ direction where Ho is surrounded by 6 oxygen atoms forming a distorted octahedron. The magnetic lattice consists of zig-zag ladders which extend along $\bf{c}$ and form a honeycomb-like pattern in the $\bf{a-b}$ plane [Fig.~\ref{fig:1}(b)].

Polycrystalline powders of $\rm SrHo_2O_4$ were prepared by solid state synthesis using $\rm Ho_2O_3$ (99.99 \%) and $\rm SrCO_3$ (99.99\%) as starting materials. The starting powders were mixed, pressed into pellets, and heated at 900, 1000 and 1100 $^\circ$C in air, each for 10 h with intermediate grinding. 4\% extra $\rm SrCO_3$ was added to the starting materials to account for evaporation of Sr during single crystal growth. After grinding and phase identification, synthesized powders were compacted into a rod using a hydraulic press. The feed rods were then sintered at 1200 $^\circ$C for 10 h in air. Single crystals with diameter 4-5 mm and length up to 60 mm were grown from the polycrystalline feed rods in a four-mirror optical floating zone furnace (Crystal Systems Inc. FZ-T-12000-X-VPO-PC) with 4$\times$3 kW xenon lamps. The growth rate was 4 mm/h with rotation rates of 5 rpm for the growing crystal (lower shaft) and 0 rpm for the feed rod (upper shaft) in a 4 bar purified argon atmosphere. Growth under oxygen containing atmospheres led to significant evaporation of phases containing Sr, and $\rm Ho_2O_3$ was formed as second phase inclusions in the as grown crystals.

\subsection{Thermomagnetic and neutron scattering measurements}

The thermomagnetic properties of $\rm SrHo_2O_4$ were measured using a Physical Properties Measurement System from Quantum Design, Inc.~with a dilution refrigerator option for measurements below $1.8$~K. Heat capacity measurements were performed using the quasi-adiabatic heat-pulse technique on a thin plate of polished single crystalline $\rm SrHo_2O_4$. Temperature dependent magnetization measurements between 2 K and 300 K along three crystalline axes were carried out using a vibrating sample magnetometer option in a magnetic field of 200 Oe.

Elastic neutron scattering (ENS) maps for $T$ down to 1.5 K were measured on MACS\cite{MACS2008} at the NIST Center for Neutron Research with $E_i=E_f=5$~meV neutrons. A 4.3 g single crystal was mounted for consecutive experiments in the $(HK0)$ and $(0KL)$ planes. Single crystal ENS measurement for $T$ down to 0.28 K in a $^3$He insert were conducted on the HB-1A instrument at the High Flux Isotope Reactor of ORNL. Samples cut to the shape of small cubes with masses of 0.34 g and 0.19 g to reduce the effects of neutron absorption were used for measurements in the $(0KL)$ and $(H0L)$ planes respectively. The sample mount was made from oxygen-free copper to ensure good thermal contact at low $T$. Temperature dependent measurements were carried out upon warming after cooling to the $0.28$~K base temperature of the pumped $^3$He system.

\section{Experimental Results}

\subsection{Thermomagnetic measurements}

\begin{figure}[t]
    \includegraphics[width=3.2 in]{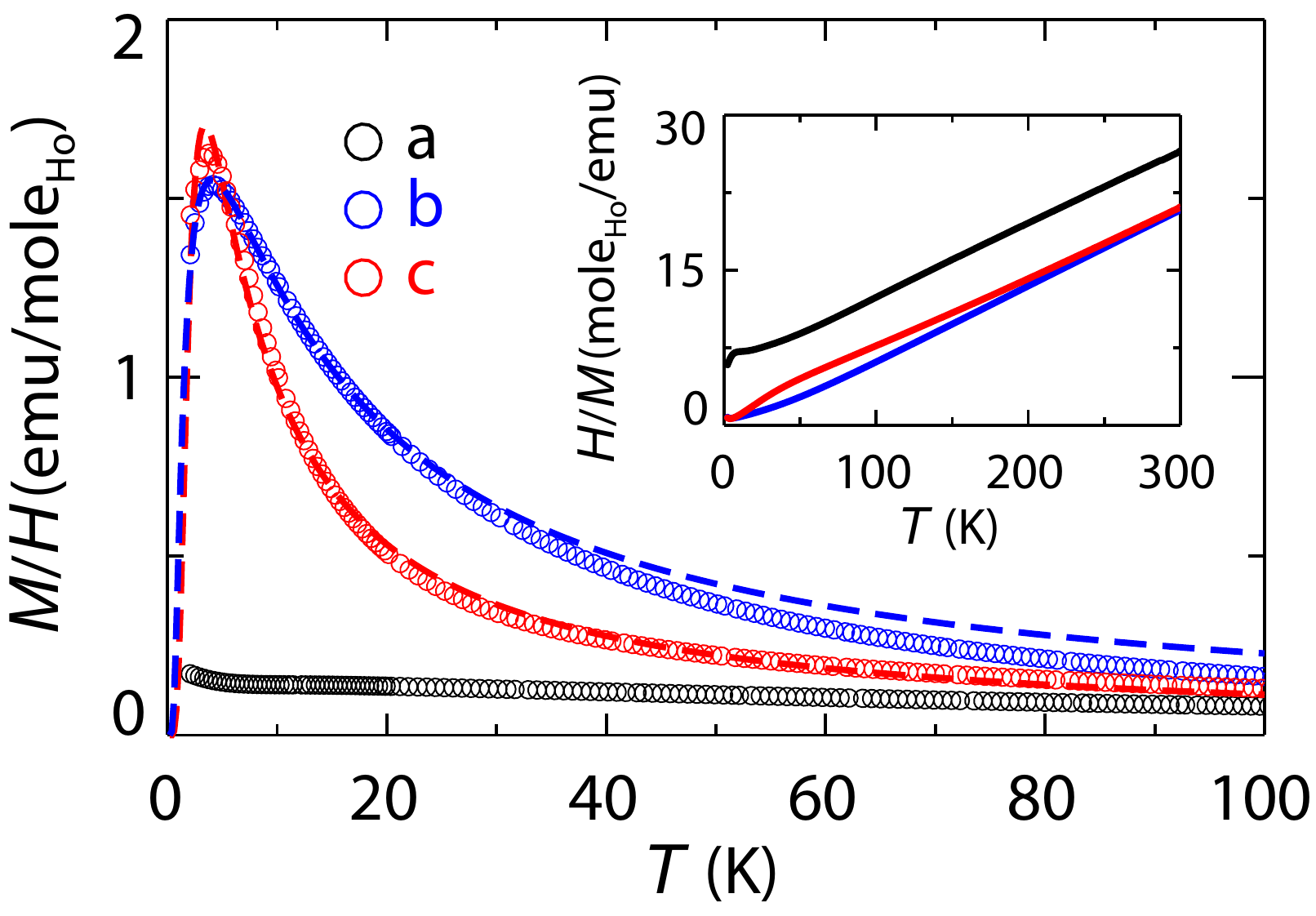}%
    \caption{\label{fig:2} Magnetic susceptibility of SrHo$_2$O$_4$ along three axes measured under 200 Oe. Dashed lines show fits to $J_1-J_2$ Ising models. The inset shows the inverse magnetic susceptibility up to 300 K. }
 \end{figure}

The magnetic susceptibility $\chi$ of $\rm SrHo_2O_4$, approximated by $M/H$ with a measuring field of $H=200$~Oe, is shown in Fig.~\ref{fig:2}. $\chi_a$ is found to be an order of magnitude smaller than $\chi_b$ and $\chi_c$, indicating strong magnetic anisotropy with a hard axis along $\bf{a}$. $\chi_b$ and $\chi_c$ increase upon cooling and form broad peaks for $T\sim 5$~K that are characteristic of antiferromagnetic (AFM) SRO.\cite{Karunadasa2005,Ghosh2011,Hayes2012} The inset in Fig.~\ref{fig:2} shows the inverse magnetic susceptibility $\chi^{-1}$ for $T$ between $2$~K and $300$~K. A Curie-Weiss analysis of the approximately linear regime between $100$~K and $300$~K results in effective moment sizes $P_{\rm eff}$ of $10.1(1)~\mu_{\rm B}$, $10.5(1)~\mu_{\rm B}$, and $11.8(1)~\mu_{\rm B}$, and Weiss temperature $\Theta_{\rm CW}$ of $-63.5(3)$~K, $16.4(1)$~K, and $-27.6(2)$~K for the $\bf{a}$, $\bf{b}$, and $\bf{c}$ directions respectively. These $P_{\rm eff}$ values are consistent with
an $^5I_8$ electronic configuration for $4f^{10}$ $\rm Ho^{3+}$. While $\Theta_{\rm CW}$ is usually associated with inter-spin interactions, such an interpretation is not straight forward here since the high temperature $\chi$ is also affected by the crystalline electric field (CEF)\cite{jensen1991} level scheme for $\rm Ho^{3+}$ in $\rm SrHo_2O_4$.

\begin{figure}[t]
    \includegraphics[width=3.4 in]{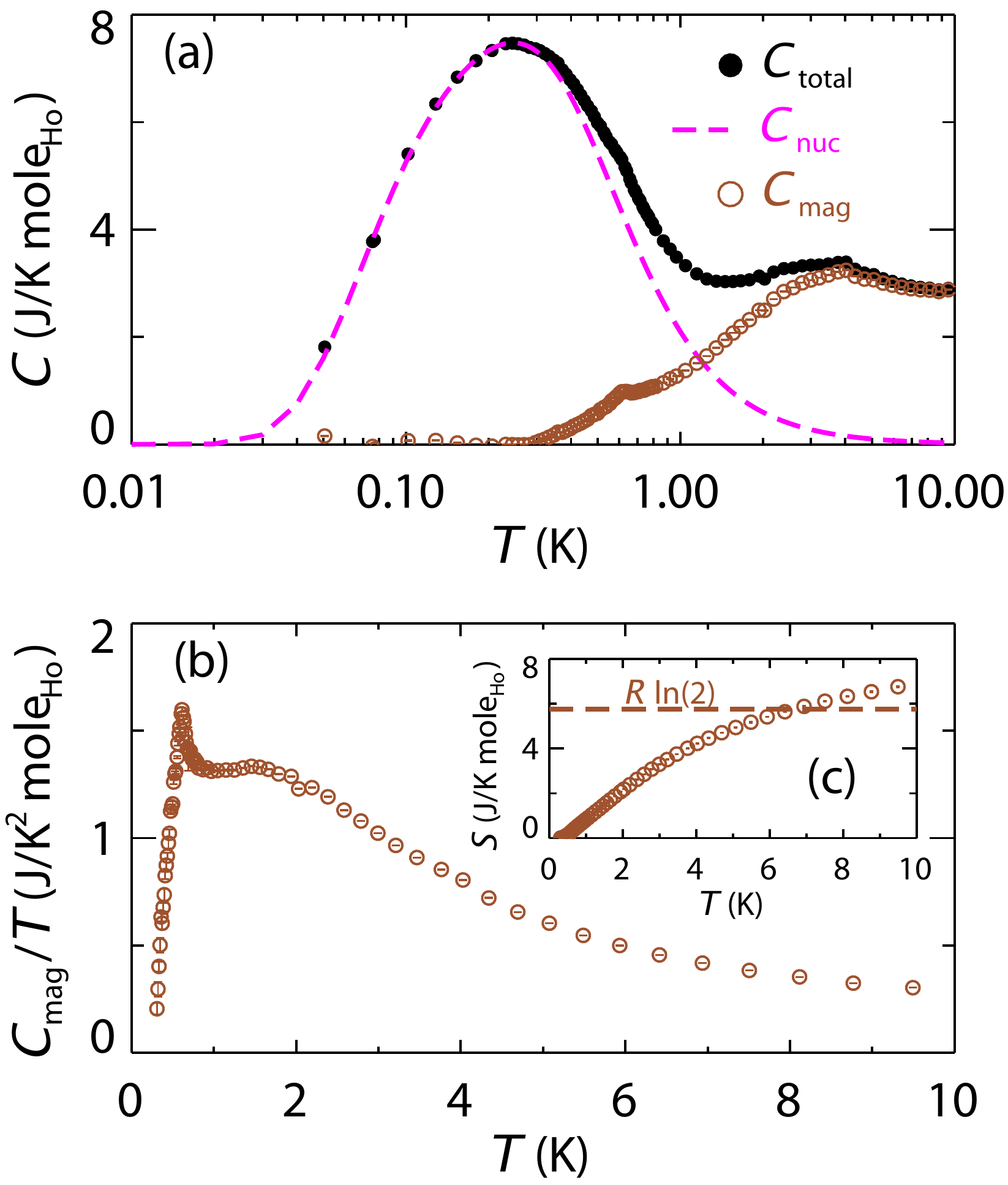}%
    \caption{\label{fig:3}(a) Specific heat of $\rm SrHo_2O_4$ as a function of $T$. Filled symbols show the experimental total specific heat; the magenta long dashed line shows the calculated specific heat due to a nuclear Schottky anomaly; open symbols show the magnetic specific heat obtained by subtracting the nuclear Schottky anomaly from the total specific heat. (b) Magnetic specific heat over $T$ versus $T$. (c) shows the entropy versus $T$. The dashed line shows the entropy of an Ising doublet.}
 \end{figure}

The temperature dependent specific heat of $\rm SrHo_2O_4$ is shown in Fig.~\ref{fig:3}(a). Filled symbols show the experimental specific heat $C_{\rm total}$. The broad peak for $T\sim$ 5 K corresponds to SRO also indicated by a maximum in $\chi$. A second anomaly is observed at $T\sim$ 0.2 K. There is a single stable isotope of holmium, $^{165}\rm Ho$ with a finite nuclear spin $I=7/2$, the multiplet of which can be split through hyperfine interactions with electronic spins. This produces a peak in the specific heat at low temperatures, the so called nuclear Schottky anomaly ($C_{\rm nuc}$).\cite{tari2003} By assuming a 7.7(2) $\mu_{\rm B}$ static magnetic moment for electronic spins of $\rm Ho^{3+}$, the low temperature $C_{\rm total}$ peak can be very well accounted for by the nuclear Schottky anomaly, as shown by the magenta long dashed line in Fig.~\ref{fig:3}(a). After subtracting $C_{\rm nuc}$ from $C_{\rm total}$, the contribution to specific heat from electronic spins ($C_{\rm mag}$) is obtained and shown as open symbols in Fig.~\ref{fig:3}(a). The phonon contribution to the specific heat is negligible in the temperature range probed.

A kink in $C_{\rm mag}$ is observed at $T_{\mathrm{N}}=0.61(2)$~K, which is even more apparent as a sharp peak in $C_{\rm mag}/T$ indicating a bulk phase transition [Fig.~\ref{fig:3}(b)].  The magnetic entropy ($S$) inferred from the area under the $C_{\rm mag}/T$ curve is shown in Fig.~\ref{fig:3}(c).

\subsection{Neutron scattering}

\begin{figure}[t]
    \includegraphics[width=3.2 in]{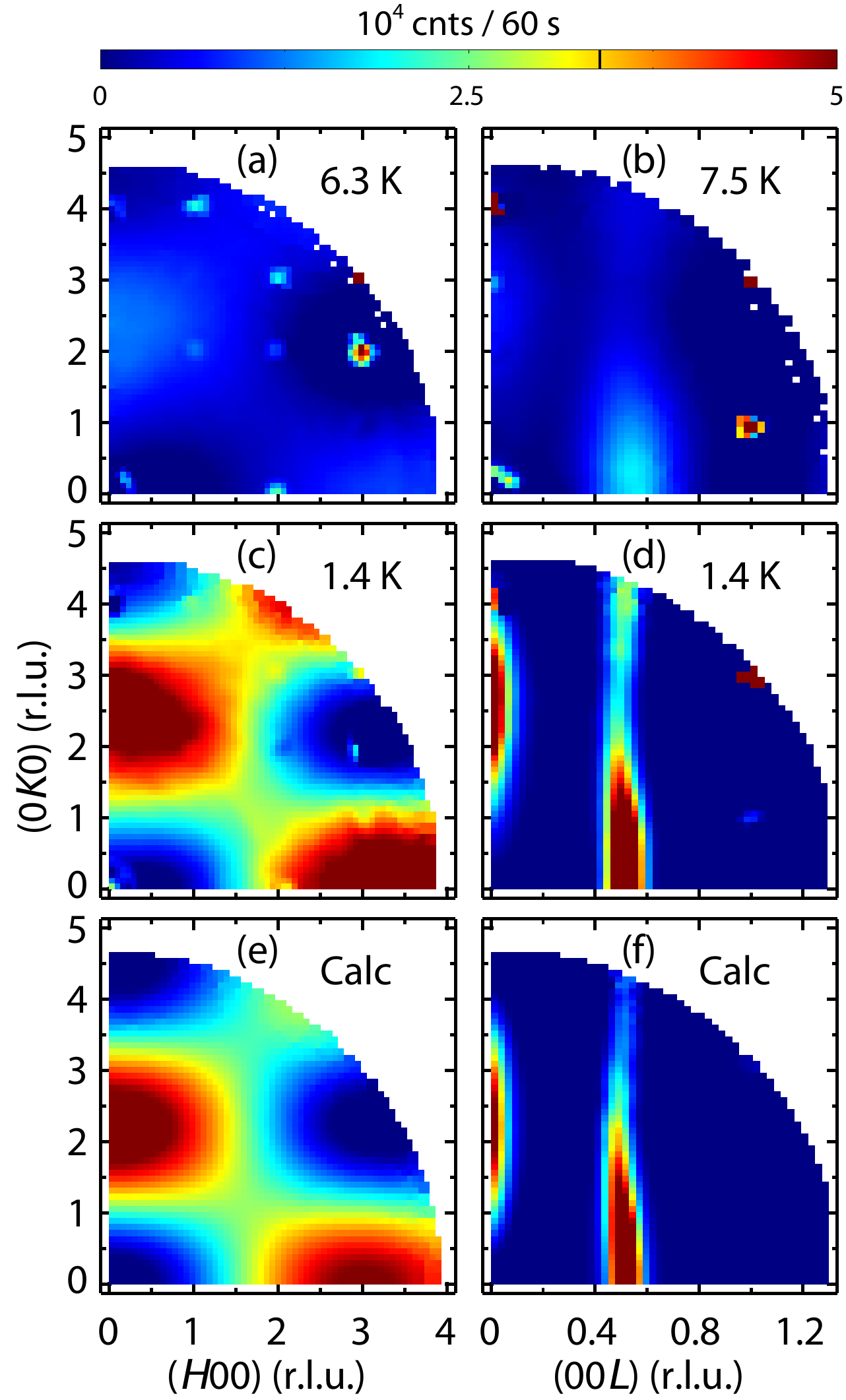}%
    \caption{\label{fig:4} $T-$dependent elastic magnetic neutron scattering indicating quasi-one-dimensional short range order in $\rm SrHo_2O_4$. (a)(c) show measurements in the $(HK0)$ reciprocal lattice plane while (b)(d) are from the $(0KL)$ plane. Measurements at 50 K were subtracted to eliminate nuclear scattering. (e) and (f) show $J_1-J_2$ model calculations at 1.4 K with exchange constants determined from Fig.~\ref{fig:2}.}
\end{figure}

For a direct view of the short range spin correlations indicated by the thermodynamic anomalies for $T\sim$ 5~K, Fig.~\ref{fig:4}(a)-(d) show ENS intensity maps covering the $(HK0)$ and $(0KL)$ planes. The strongly anisotropic nature of the wave vector dependence indicates quasi-1D correlations along the $\bf{c}$-axis consistent with previous data.\cite{Ghosh2011,Young2013} The modulation in the $(HK0)$ plane takes the form of a checker-board-like structure [Fig.~\ref{fig:4}(a) and (c)] and is associated with intra-ladder correlations. The fact that this scattering occurs for $\bm{q_c}$ $=0$ but vanishes near $\bm{q}$ $=0$ indicates an AFM structure that is not modulated along $\bf{c}$. The $(0KL)$ intensity map reveals another type of correlations with $\bm{q_c}$ $\sim 0.5\bf{c^*}$, where $\rm c^*\equiv \frac{2\pi}{c}$. The intensity maximum near $\bm{q_c}$ $\sim 0.5\bf{c^*}$ indicates spins displaced by $\bf{c}$ are anti-parallel. As will be discussed in Sec.~\ref{sec:CEF}, the single ion magnetic anisotropy in $\rm SrHo_2O_4$ allows an unambiguous association of red sites with $\bm{q_c}$ $=0$ correlations while spins on blue sites host $\bm{q_c}$ $\sim 0.5\bf{c^*}$ type correlations. For clarity this assignment will be employed from this point though it will not be justified until Sec.~\ref{sec:CEF}. The correlation lengths along $\bf{c}$ can be estimated by 2/(FWHM$_{\rm expt}^2$-FWHM$_{\rm reso}^2$)$^{-\frac{1}{2}}$, where FWHM$_{\rm expt}$, FWHM$_{\rm reso}$ are experimental and instrumental full width at half maximum. At $T=1.5$~K we find the correlation lengths along $\bf{c}$ are indistinguishable at 13.3(3)\AA\ and 13.6(3) \AA\ for red and blue ladders respectively.

\begin{figure}[t]
    \includegraphics[width=3.2 in]{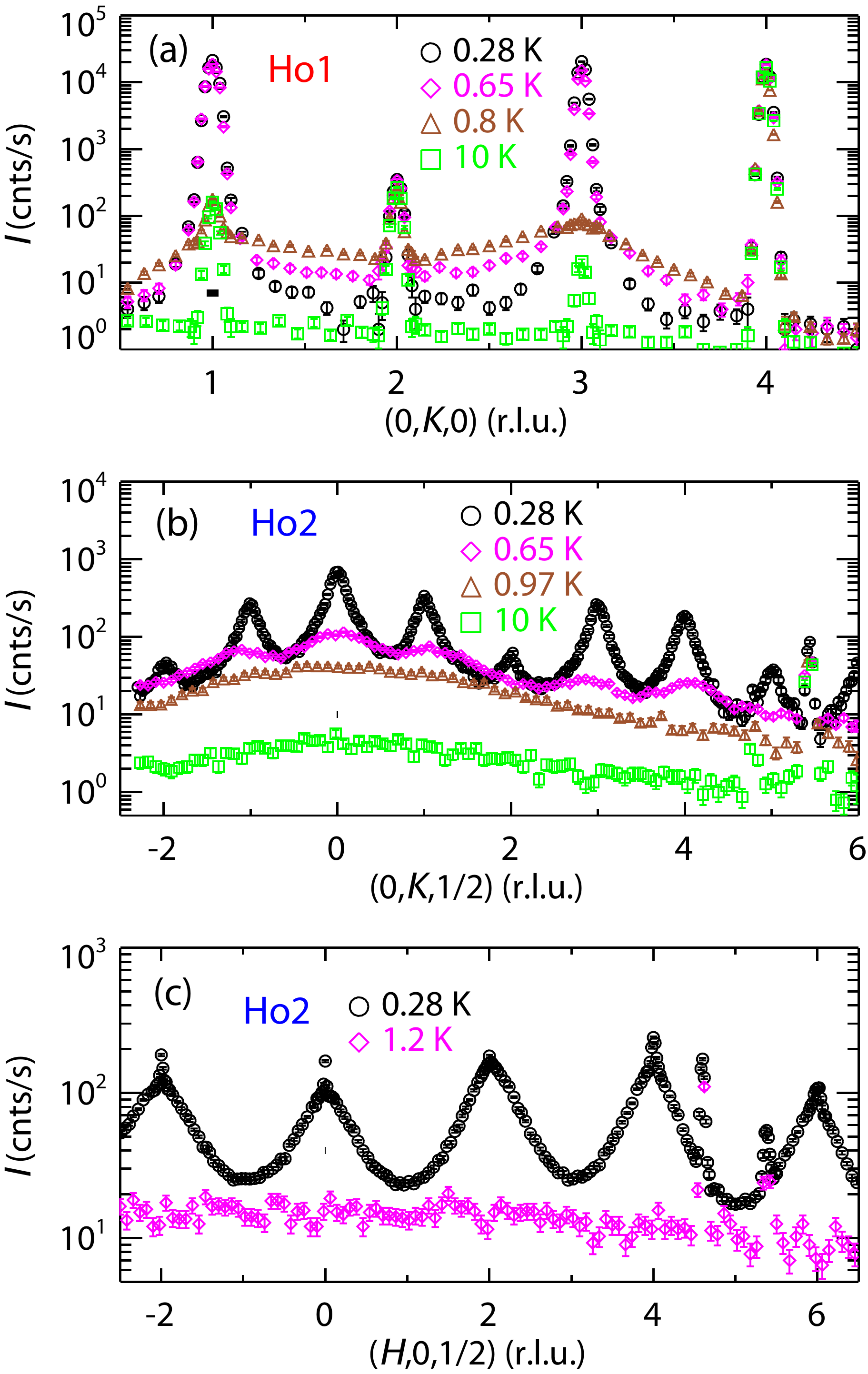}%
    \caption{\label{fig:5}Thermal evolution of inter-chain correlations for red and blue sites probed by neutrons. (a) shows $(0,K,0)$ scans probing the correlations along $\bf{b}$ between red chains at several temperatures. (b) and (c) respectively show $(0,K,\frac{1}{2})$ and $(H,0,\frac{1}{2})$ scans that probe correlations along $\bf{b}$ and $\bf{a}$ for blue chains. Horizontal bars represent instrumental resolution. Sharp peaks at $K \sim 5.5$ in (b), at $H \sim 4.6$ and $H \sim 5.4$ in (c) are the results of Bragg scattering from the copper sample mount. }
\end{figure}

The evolution of these two kinds of short range correlations across $T_\mathrm{N}$ were probed by single crystal ENS in the $(0KL)$ and $(H0L)$ planes down to $T \sim 0.3$~K. Firstly $L-$ scans at all accessible $(\mathrm{n}_a,\mathrm{n}_b,\mathrm{n}_c)$ and $(\mathrm{n}_a,\mathrm{n}_b,\mathrm{n}_c+1/2)$ magnetic peaks ($\mathrm{n}_a, \mathrm{n}_b, \mathrm{n}_c$ being integers) are resolution limited at $T=0.28$~K. This is evidence of quasi-1D correlations over length scales exceeding 286(5) \AA\ and 100(1) \AA\ respectively for red and blue ladders. These lower bounds were obtained from $L-$scans at $(030)$ and $(00\frac{1}{2})$ respectively. The different limits arise because the $L-$scan is a rocking scan for red ladders but a longitudinal scan for blue ladders, which results in better resolution for red ladders.

Temperature dependent $(0K0)$ scans for $K\in [0.5,4.5]$ between 0.28 K and 10 K were acquired to probe correlations between red chains with relative displacement along $\bf{b}$ [Fig.~\ref{fig:5}(a)]. Upon cooling, a broad intensity modulation develops and turns into resolution limited Bragg peaks for $T$ below $T_\mathrm{N}=0.68(2)$~K. Measurements at $(200)$ and $(030)$ indicate the correlation length for red spins exceeds 57(1) \AA\ and 64(1) \AA\ respectively along $\bf a$ and $\bf b$ at $T=0.28$~K.

\begin{figure}[t]
    \includegraphics[width=3.2 in]{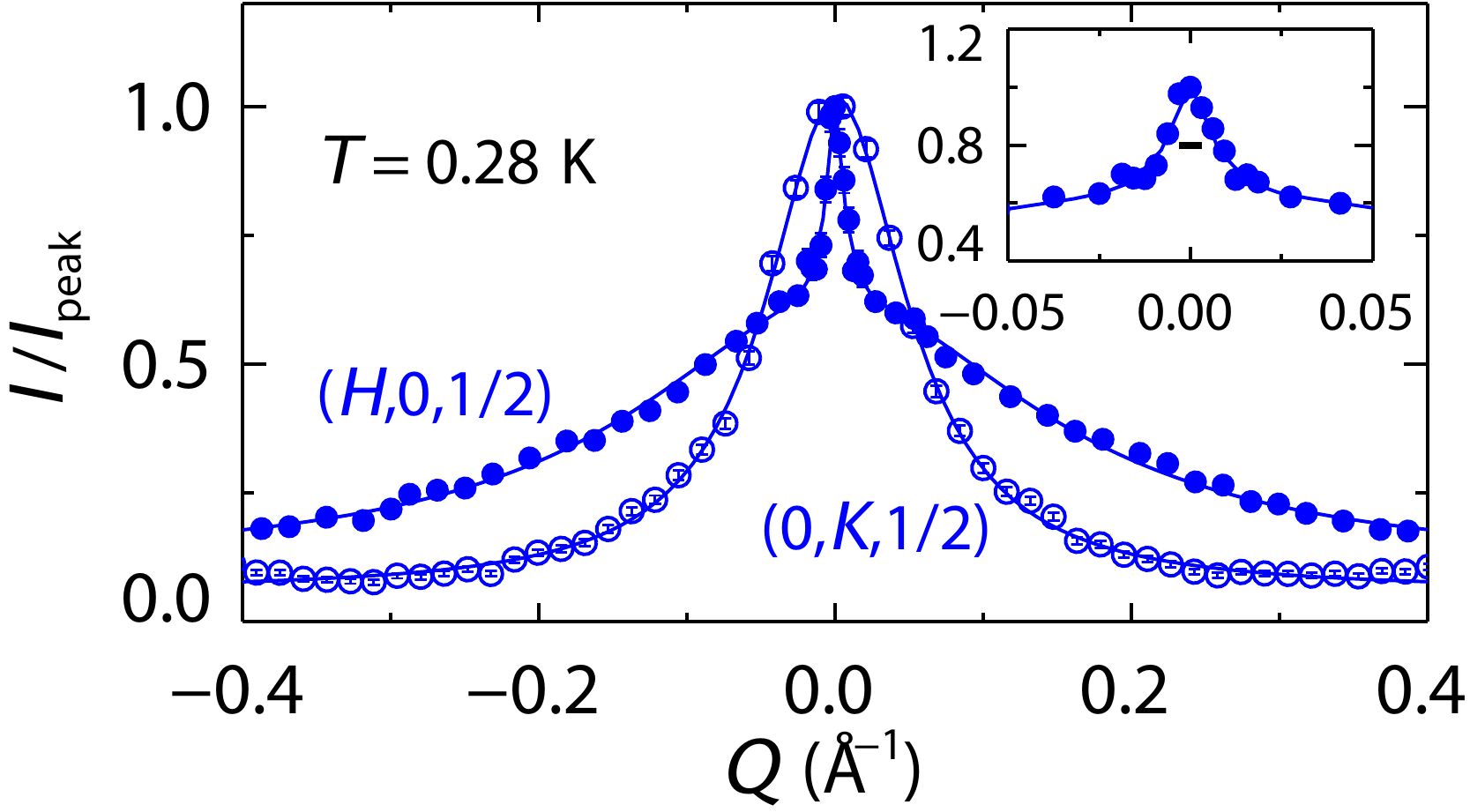}%
    \caption{\label{fig:6}Inter-chain correlations for blue sites in $\rm SrHo_2O_4$ probed by neutrons. Filled and empty circles are from $H-$ and $K-$scans at $(00\frac{1}{2})$ measured at 0.28 K. The solid lines show fits to the data. Insert shows a sharp component of $H-$scan in detail, the horizontal bar is the instrumental resolution. Data from different sample orientations are normalized to the peak count rate.}
\end{figure}

Despite a spin correlation length exceeding 100(1) \AA\ along $\bf c$, blue chains however, fail to develop conventional long range inter-chain correlations [Fig.~\ref{fig:5}(b)(c)]. While these inter-chain correlations are enhanced at low $T$ as manifested by sharpening of the peaks, the peak width remains much broader than the instrumental resolution shown by horizontal bars. A detailed view of $H-$ and $K-$scans through $(00\frac{1}{2})$ at $T=0.28$~K is provided in Fig.~\ref{fig:6}. The broad modulations are well described by Lorentzian fits that correspond to correlation length along $\bf{a}$ and $\bf{b}$ of just 6.0(1) \AA\ and 17.5(3) \AA\ respectively. The $H-$scan however, also includes a curious small sharp component that indicates some correlations between blue chains persists to a separation of 165(9) \AA\ perpendicular to their easy axis. This observation merits further theoretical and experimental exploration.

\section{Analysis}

\subsection{Single ion anisotropy}
\label{sec:CEF}

\begin{figure}[t]
    \includegraphics[width=3.4 in]{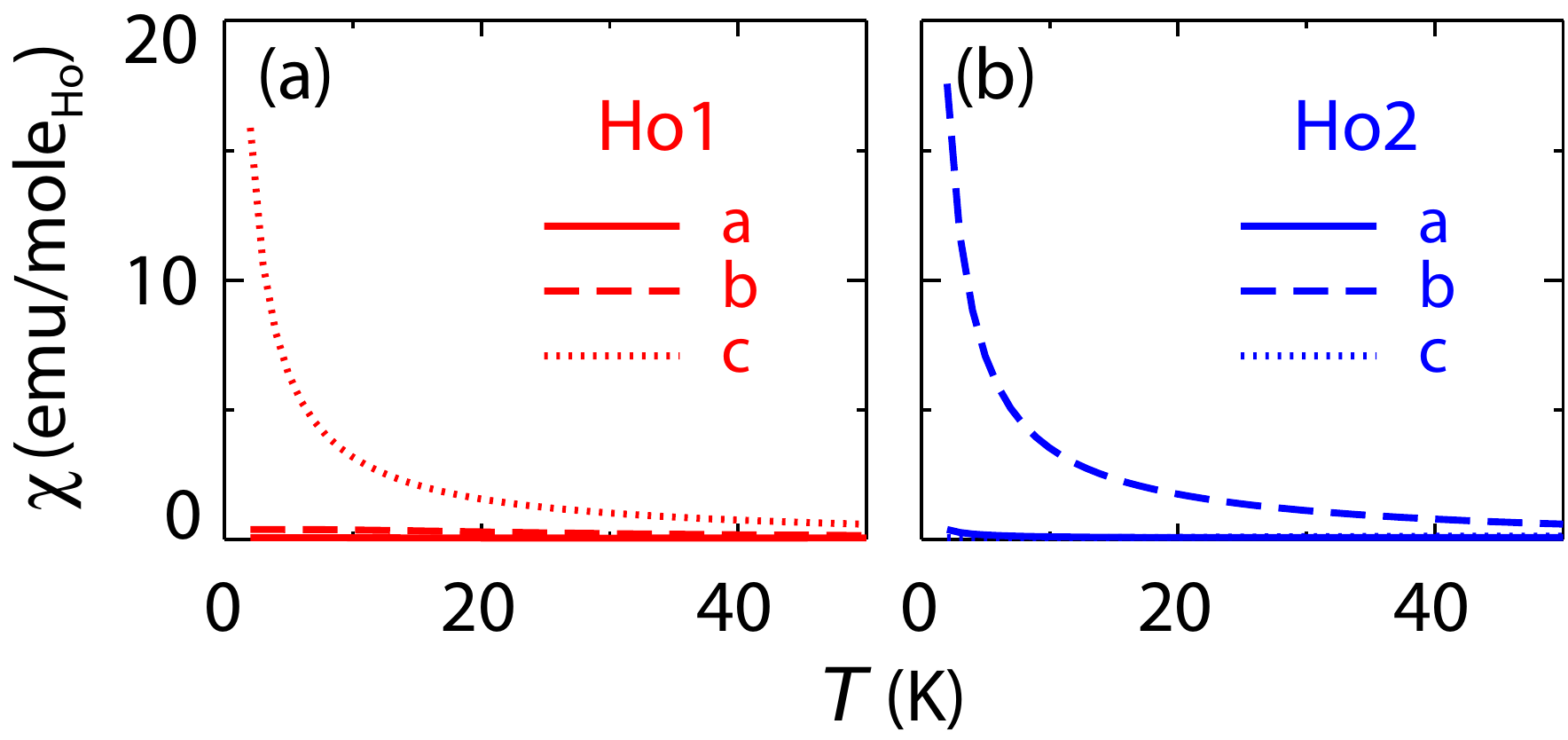}%
    \caption{\label{fig:10}Calculated single ion magnetic susceptibility for (a) red Ho1 site and (b) blue Ho2 site in $\rm SrHo_2O_4$ along three axes based on CEF level schemes calculated according to point charge approximation.}
\end{figure}

As has been suggested in previous sections, magnetic anisotropy plays an all important role in the magnetic properties of $\rm SrHo_2O_4$. In the anisotropic environment of the solid, the $J=8$ multiplet of Ho$^{3+}$ is split into multiple levels resulting in the anisotropic susceptibility\cite{jensen1991} shown in Fig.~\ref{fig:2}. The $C_{s}$ point group symmetry of both red and blue $\rm Ho^{3+}$ sites implies an easy magnetic axis either along $\bf{c}$ or within the $\bf{ab}$ plane. A CEF calculation based on the point charge approximation\cite{Hutchings1964}, where CEF at Ho site is approximated by the static electric field from 6 surrounding $\rm O^{2-}$ ions, shows both red and blue sites have a doublet ground state. The single ion magnetic susceptibility calculated based on the point charge CEF level schemes (Fig.~\ref{fig:10}) reveals the Ising doublet ground state for red Ho1 (blue Ho2) sites have an easy axis along $\bf{c}$ ($\bf{b}$). We infer the temperature dependent susceptibility (Fig.~\ref{fig:2}) for fields along $\bf{c}$ and $\bf{b}$ are due to red and blue sites respectively. For blue sites a finite moment along $\bf{a}$ is allowed by symmetry, however, since $\chi_a$ is minimal it will be neglected.

With this information about the magnetic anisotropy for the two Ho sites, it can be inferred that the scattering in the $(HK0)$ plane in Fig.~\ref{fig:4}(a)(c) is due to red ladders because there is no decrease in intensity for $\bm{q} \parallel$ $(0K0)$ which would be the case for easy $\bf{b}$-axis blue sites. This is a consequence of the polarization factor in magnetic neutron scattering which ensures that neutrons only probe magnetic moments perpendicular to $\bm{q}$.\cite{squires2012} An analogous polarization argument shows that the scattering in $(0KL)$ plane with $\bm{q_c}$ $\sim 0.5\bf{c^*}$ arises from blue sites.

\subsection{ANNNI model}

 \begin{table}[t]
 \caption{\label{tab:1}Magnitude of dipolar interaction energies between neighboring spins in $\rm SrHo_2O_4$ assuming Ising moment size of 6.2 $\mu_{\rm B}$ and 9.9 $\mu_{\rm B}$ on red and blue sites respectively. The numbering of $\rm Ho^{3+}$ ions is as shown in Fig.~\ref{fig:1}(b). Intra-ladder dipolar energies (row 1-4) are positive/negative for FM/AFM interactions. For comparison, the corresponding ANNNI model exchange constants inferred from susceptibility fits are in the third column.}
 \begin{ruledtabular}
 \begin{tabular}{l c r}
  Pair of Spins & Dipolar Energy (meV) & $J_{\rm ANNNI}$ (meV) \\
  \textcolor{black}{Ho1(1)} - \textcolor{black}{Ho1(3)} & -0.01 & \textcolor{black}{$J_{r1}$} -0.10(2) \\
  \textcolor{black}{Ho1(1)} - \textcolor{black}{Ho1(1)}+\bf{c} & ~0.10 & \textcolor{black}{$J_{r2}$}~~0.04(3)  \\
  \textcolor{black}{Ho2(1)} - \textcolor{black}{Ho2(3)} & ~0.08 & \textcolor{black}{$J_{b1}$} -0.14(3) \\
  \textcolor{black}{Ho2(1)} - \textcolor{black}{Ho2(1)}+\bf{c} & -0.13 & \textcolor{black}{$J_{b2}$} -0.21(1)\\
  \textcolor{black}{Ho1(1)} - \textcolor{black}{Ho2(3)} & ~0.05 &\\
  \textcolor{black}{Ho1(4)} - \textcolor{black}{Ho2(1)} & ~0.00 &\\

 \end{tabular}
 \end{ruledtabular}
 \end{table}

The dominant intra-ladder SRO (Fig.~\ref{fig:4}) suggests the minimal model for $\rm SrHo_2O_4$ is a collection of two types of independent Ising zig-zag spin ladders with different inter-leg interactions $J_{r1},J_{b1}$ and intra-leg interactions $J_{r2},J_{b2}$ for the red and blue chains respectively [Fig.~\ref{fig:1}(b)]. The underlying model for both chains is the exactly solvable 1D ANNNI model\cite{Selke1988213}:
\begin{eqnarray} \label{eq:1}
 H=\sum_{i}-J_1S_iS_{i+1}-J_2S_iS_{i+2},
\end{eqnarray}
where $S_i=\pm 1$. Dipolar interactions between spins in $\rm SrHo_2O_4$ can be comparable to $T_{\rm N}$ because of the large magnetic moment size for $\rm Ho^{3+}$ ions. To be specific, consider magnetic moment size of $6.2~\mu_{\rm B}$ and $9.9~\mu_{\rm B}$ oriented along the easy axes of red and blue $\rm Ho^{3+}$ sites respectively. These are the magnetic moment sizes for the two Ho sites estimated in this study that will be justified in detail in Sec.~\ref{sec:momentsize}. The corresponding dipolar interaction energies between neighboring $\rm Ho^{3+}$ ions are listed in Table.~\ref{tab:1}.  These energies are found to be of order a Kelvin and extend with considerable strength to further neighbors. The ANNNI model should thus be considered as a minimal effective model to describe each of the spin ladders in $\rm SrHo_2O_4$. Antisymmetric Dzyaloshinskii-Moriya (DM) interactions are also allowed in the low crystalline symmetry of $\rm SrHo_2O_4$. The strong Ising anisotropy however, extinguishes intra-ladder DM interactions because all spins within each type of ladder are oriented along the same easy axis. While DM interactions between red and blue ladders are allowed, the different modulation wave vectors for the two types of ladders (0 and $0.5\bf{c^*}$ respectively) render these and all other inter-ladder interactions ineffective at the mean field level. This may explain why the simple model of independent ANNNI chains that we shall explore in the following provides a good basis for describing the magnetism of $\rm SrHo_2O_4$ outside of the critical regime near $T_{\rm N}$.

To determine the exchange constants $J_1$ and $J_2$ we fit the anisotropic susceptibility to the analytical result for the susceptibility of the ANNNI model. The exchange constants for red and blue chains are denoted by $J_{r1,2}$ and $J_{b1,2}$  respectively. The uniform magnetic susceptibility $\chi$ can be related to the two point correlation function as follows
\begin{eqnarray}
\chi\equiv\lim_{h \to 0}\frac{\partial \langle M\sum_i S_i \rangle}{\partial h}=NM^2\beta\tilde{G}(\bm{q}=0),
\end{eqnarray}
where $N$ is the number of sites in the spin chain, $M$ is the dipole moment of each spin, $h$ is the magnetic field, $\beta=1/k_{\mathrm{B}}T$, $\tilde{G}(\bm{q}=0)=\sum_iG(i)$, and $G(i)\equiv \braket{S_0S_i}$ is the two-point correlation function for the 1D ANNNI model. The fits were restricted to data points with $T\le20$ K where the influence from higher CEF levels can be neglected.

For red chains (accessible through $\chi_c$), the best fit is achieved for $J_{r1}=-0.10(2)$ meV and $J_{r2}=0.04(3)$ meV. The corresponding calculated susceptibility is shown as a red dashed line in Fig.~\ref{fig:2}. These exchange parameters define an unfrustrated ANNNI chain where all interactions are simultaneously satisfied by the N\'{e}el structure ($\uparrow\downarrow\uparrow\downarrow$). $\chi_b$ for blue chains is best fit with $J_{b1}=-0.14(3)$ meV and $J_{b2}=-0.21(1)$ meV. While these competing interactions produce incommensurate short range correlations at finite temperatures, the ground state is the double N\'{e}el structure ($\uparrow\uparrow\downarrow\downarrow$).\cite{Selke1988213} The magnetic moment sizes extracted from fitting the susceptibility data for red and blue sites are 5.5(3) $\mu_{\rm B}$ and 8.1(2) $\mu_{\rm B}$ respectively, which are consistent with neutron diffraction measurements.\cite{Poole2014,Young2012} These effective Ising exchange constants are compared to the dipolar interaction strengths in Table~\ref{tab:1}. The significant discrepancies might be accounted for by contributions to the effective ANNNI interactions from superexchange, longer range dipole interactions, as well as higher order  effects from inter-ladder interactions.

A critical test of the quasi-1D ANNNI model for $\rm SrHo_2O_4$  is offered by Fig.~\ref{fig:4}. Frames (e) and (f) show a calculation of the diffuse magnetic neutron scattering intensity at the given temperature from such spin chains based on the Fourier transformation of the two-point correlation function $G(r)$ for the exchange constants derived from $\chi_c$ and $\chi_b$ and the particular crystalline structure of the ladders. Only an overall scale factor and a constant background were varied to achieve the excellent account of the ENS data in Fig.~\ref{fig:4}(c)-(d). Though no correlations between spin chains are included, the finite width of the zig-zag spin ladders and the two different red ladder orientations in $\rm SrHo_2O_4$ [Fig.~\ref{fig:1}(b)] produce the checker-board-like structure in the $(HK0)$ plane. It is remarkable that a purely 1D model can account for the magnetism of a dense 3D assembly of spin chains. Contributing to this are surely the different spin orientations for red and blue sites and the incompatible modulation wave vectors.

\subsection{Low temperature magnetic structure}
\label{sec:momentsize}

\begin{figure}[t]
    \includegraphics[width=3.2 in]{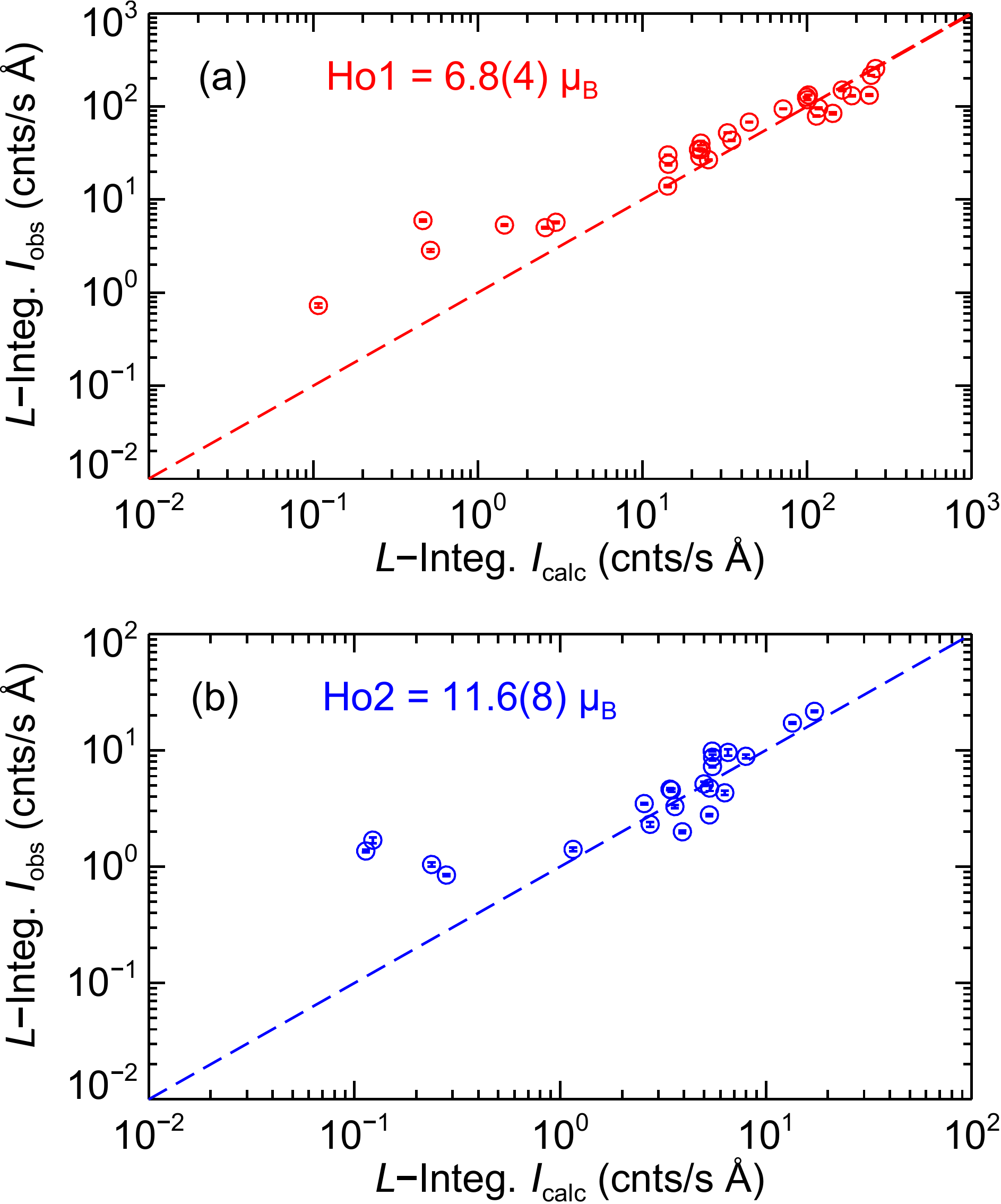}%
    \caption{\label{fig:7}Magnetic structure refinement for $\rm SrHo_2O_4$ based on $L-$integrated peak intensities measured for $T= 0.3$~K. (a) and (b) show the refinement results for red and blue sites respectively. Long dashed lines represent the $y=x$ line.}
\end{figure}

While the strictly 1D ANNNI model can only form LRO at $T=0$ K, finite inter-chain interactions can induce LRO at finite $T_{\rm N}$. ENS (Fig.~\ref{fig:5}) reveal that the red sites form 3D LRO while the blue sites develop long range correlations only along the chain direction. To characterize the corresponding magnetic structures and to estimate the static magnetic moment sizes, a magnetic structure refinement was carried out based on ENS measured for $T=0.3$~K.

We first consider the 3D ordered red sites. Since the propagation vector $\bm{q_c}=0$, magnetic Bragg scattering overlaps with nuclear Bragg peaks. For unpolarized neutrons employed in this study, the total Bragg scattering intensity is simply the sum of these two contributions.

Representation analysis\cite{Wills2000680} shows that for $\bm{q_c}=0$ there are two magnetic structures that are compatible with both the easy $\bf{c}$-axis anisotropy and the short range magnetic order inferred from the diffuse neutron scattering pattern (Fig.~\ref{fig:4}). Using the labeling of red Ho1 atoms in the unit cell as shown in Fig.~\ref{fig:1}(b), these two structures can be represented as $\phi_{1,2}\equiv (\bm{m}_1,\bm{m}_2,\bm{m}_3,\bm{m}_4)=\mathrm{m}\bm{\hat{c}}(1,\pm1,-1,\mp1)$, where $\bm{m}_i$ is the magnetic moment on atom $i$, $\mathrm{m}$ is the moment size, and $\bm{\hat{c}}$ is a unit vector along $\textbf{c}$-axis. The experimental observation that all nuclear forbidden Bragg peaks in the $(H0L)$ plane with even $H$ indices have negligible magnetic scattering intensity shows that $\phi_1$ is the appropriate magnetic structure, since $\phi_2$ would give rise to magnetic Bragg peaks at such locations.

To extract the magnetic moment size for Ho1, the $L-$integrated intensities for all accessible Bragg peaks with integer indices, which contain both nuclear and magnetic scattering contributions, were compared to the calculated neutron diffraction intensity for the $\phi_1$ structure. The nuclear structure factors were calculated according to the crystal structure determined in a previous study.\cite{Karunadasa2005} Measurements in the $(0KL)$ and $(H0L)$ planes were co-refined while keeping the ratio of scale factors in the two reciprocal lattice planes fixed at the mass ratio for the samples employed in each reciprocal lattice plane. The best fit shown in Fig.~\ref{fig:7}(a) yields a moment size of $6.8(4)~\mu_\mathrm{B}$ for the red Ho1 sites. The correspondence with the red chain moment of 5.5(3) $\mu_\mathrm{B}$ derived from ANNNI fits to $\chi_c$ corroborates the assumptions that underlie this analysis.

For the blue sites, where the correlation length is short along $\bf{a}$ and $\bf{b}$ directions, we approximate the peak shape as the product of a sharp Gaussian along $\bf{c}$ with two broad Lorentzians along $\bf{a}$ and $\bf{b}$. For measurements in the $(0KL)$ plane, the peak width of the Gaussian and the in-plane Lorentzian were obtained by fitting the experimental data, while the peak width for the out of plane Lorentzian was assumed to be identical at all peak positions and was approximated by the average peak width along $\bf{a}$ measured in the $(H0L)$ plane. Note that when taking the average, each peak width was weighted by the corresponding integrated intensity so that stronger peaks contribute with larger weight to the average: $\overline{\mathrm{FWHM}}=\sum_i \mathrm{FWHM}_i\cdot I_i/\sum_i I_i$. Following the same procedure, the out of plane Lorentzian width for peaks within the $(H0L)$ plane were obtained from measurements in the $(0KL)$ plane.

Representation analysis for $\bm{q_c}=0.5\bf{c^*}$ allows spin structures of the form $\psi=\mathrm{m}_1\bm{\hat{b}}(1,1,0,0)+\mathrm{m}_2\bm{\hat{b}}(0,0,1,1)$. Since there is no 3D long range order, this magnetic structure only reflects the locally ordered pattern. It is assumed that $|\mathrm{m}_1|=|\mathrm{m}_2|$ since all four Ho2 ions in the unit cell are equivalent in the paramagnetic phase and no significant improvement in refinement was obtained by allowing $\mathrm{m}_1$ and $\mathrm{m}_2$ to vary independently. With this constraint there are still two different magnetic structures $\psi_{1,2}=\mathrm{m}\bm{\hat{b}}(1,1,\pm1,\pm1)$, which can be regarded as two domain types that are related by mirror reflection about the $\bf{ab}$ plane. With no $a~priori$ reason to favor one domain over the other, it is assumed that both domains contribute equally. Using the same scale factor obtained from refinement of the red Ho1 moment size, least squares fit to the $L-$integrated intensities of all accessible peaks of the form $(0K\frac{2n+1}{2})$ and $(H0\frac{2n+1}{2})$ was conducted. This resulted in a moment size of $11.6(8)~\mu_\mathrm{B}$ for blue Ho2. Fig.~\ref{fig:7}(b) provides a comparison of the measured and calculated integrated intensities.

The analysis of anisotropic diffuse scattering that is the basis for the moment sizes extracted for blue sites is subject to systematic uncertainties that are not reflected in the error bars. We therefore consider the magnetic moment sizes extracted from the neutron measurements [6.8(4) $\mu_{\rm B}$ and 11.6(8) $\mu_{\rm B}$ for red and blue Ho] consistent with those obtained from the ANNNI susceptibility fits [5.5(3) $\mu_{\rm B}$ and 8.1(2) $\mu_{\rm B}$ respectively]. Combining these results, which are subject to different systematic errors leads to an experimental average result of 6.2(3) $\mu_{\rm B}$ and 9.9(4) $\mu_{\rm B}$ for red and blue sites respectively. The difference in the moment sizes for red and blue sites indicates the different CEF environment for these two sites. The moment on blue sites is close to the maximal magnetic moment size of 10 $\mu_{\rm B}$ for a Ho$^{3+}$ ion, which indicates a strong Ising character for blue Ho2 ions.

\subsection{Temperature dependent spin correlations}

\begin{figure}[t]
    \includegraphics[width=3.2 in]{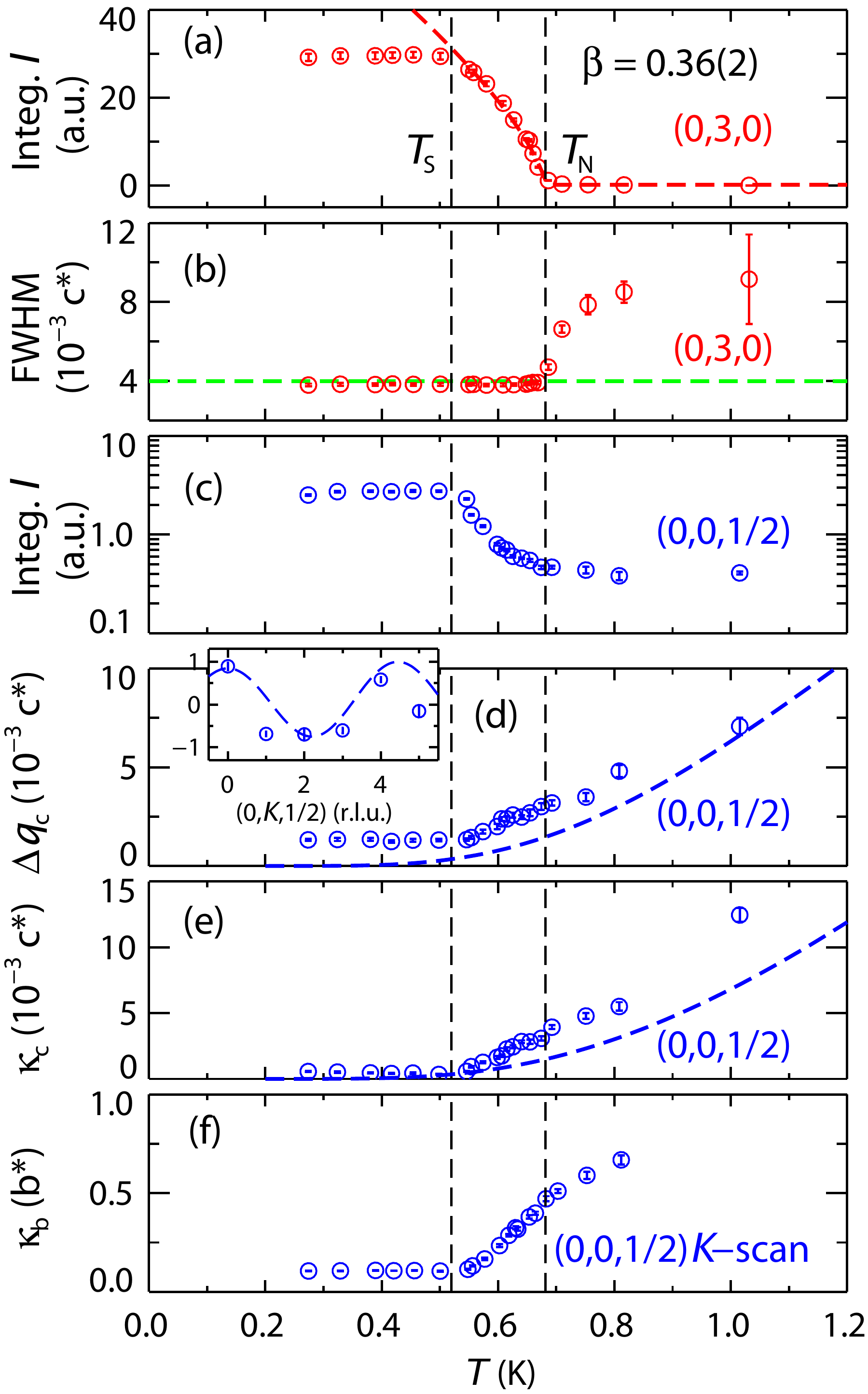}%
    \caption{\label{fig:8}Spin correlations versus $T$ in $\rm SrHo_2O_4$ probed by neutrons. (a)-(e) show results from $L-$scans at $(030)$ (red) and $(00\frac{1}{2})$ (blue) that probe correlations along red and blue chains respectively. (a) and (b) show integrated intensities and peak widths for $(030)$. The green dashed line in (b) shows the instrumental resolution. (c) shows integrated intensities for $(00\frac{1}{2})$. (d) shows the peak shift from $(00\frac{1}{2})$. The inset shows the peak shift along $\bf c^*$ from $(0K\frac{1}{2})$. The dashed line shows the predicted shift based on the $J_1-J_2$ model. (e) shows the inverse correlation length $\kappa_c$ derived as the half width at half maximum of resolution convoluted Lorentzian fits. (f) shows $\kappa_b$ extracted from $K-$scans at $(00\frac{1}{2})$. Black dashed lines indicate $T_\mathrm{N}$ and $T_\mathrm{S}$. Blue dashed lines in (d) and (e) are ANNNI model calculations based only on the exchange constants obtained from the data in Fig.~\ref{fig:2}.}
\end{figure}

To probe the interplay between the two types of spin ladders, scans for a range of $T$ were carried out through the $(030)$ and $(00\frac{1}{2})$ peaks. These peaks respectively arise from red and blue chains (Fig.~\ref{fig:8}). The result for red sites is shown in Fig.~\ref{fig:8}(a) and (b). The integrated intensity of the peak, a measure of the staggered magnetization squared, grows below $T_\mathrm{N}$ then saturates at $T_\mathrm{S}=0.52(2)$~K. This is consistent with a second order phase transition in a uniaxial spin system where the gap in the magnetic excitation spectrum produces a characteristic saturation temperature. Near $T_\mathrm{N}$, the critical exponent $\beta=0.36(2)$ is consistent with that for the 3D Ising model [$\beta_{3DI}=0.3258(14)$], but also indistinguishable from the 3D XY [$\beta_{3DXY}=0.3470(14)$] and 3D Heisenberg models [$\beta_{3DH}=0.3662(25)$].\cite{PhysRevD.60.085001} The peak width [Fig.~\ref{fig:8}(b)] decreases markedly upon cooling towards $T_\mathrm{N}$ signalling the development of commensurate long range correlations among red spins.

A rather different situation is found for the blue chains. There is no anomaly in the temperature dependent $L-$integrated intensity of the $(00\frac{1}{2})$ peak at $T_\mathrm{N}$ but a gradual increase upon cooling that terminates at $T_\mathrm{S}$ [Fig.~\ref{fig:8}(c)]. The nature of spin correlations along the blue chains is probed by the position and width of the $(00\frac{1}{2})$ peak. Both evolve continuously across $T_\mathrm{N}$ in  semi-quantitative agreement with the ANNNI model, using the  parameters that also describe the susceptibility and diffuse neutron scattering data.  The trend however, ceases at $T_\mathrm{S}$ with a peak center position of $0.501\bf{c^*}$.  The deviation, $\Delta q_c$, from the commensurate position $0.5\bf{c^*}\rm$ is significant and a long wave length modulated structure is apparent as an oscillation of the centers for other $(0K\frac{1}{2})$ type peaks with $K=1,2,3,4,5$ [inset to Fig.~\ref{fig:8}(d)]. This is consistent with $G(r)$ for the incommensurate zig-zag ladder, which is indicated by the dashed line in the inset.

\section{Discussion and Conclusion}
\label{sec:discussion}

\begin{figure}[t]
    \includegraphics[width=3.4 in]{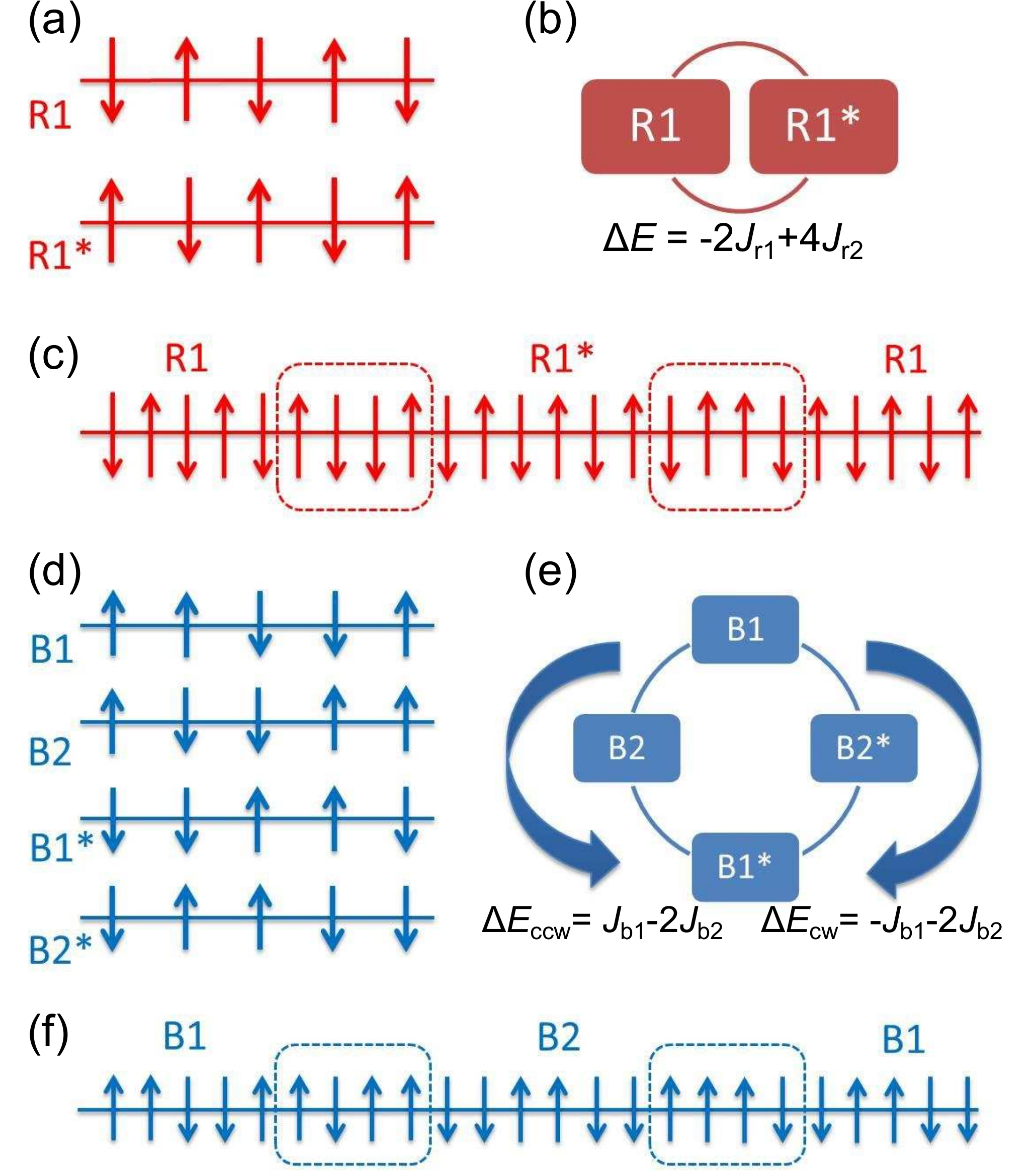}%
    \caption{\label{fig:9}Ground state degeneracy and domain walls of ANNNI chains in $\rm SrHo_2O_4$. (a) and (b) show the two fold degenerate ground states for red chains and the corresponding domain wall. (c) illustrates that a $R1$ to $R1^*$ domain wall is identical to a $R1^*$ to $R1$ domain wall (up to time reversal). (d) and (e) show the more complicated situation for blue chains where the ground state is four fold degenerate and the domain walls are chiral. (f) illustrates the chiral character of the domain walls in blue chains: a $B1$ to $B2$ domain wall is different from a $B2$ to $B1$ domain wall. Refer to Sec.~\ref{sec:discussion} for a detailed description. Dashed rectangles in (c) and (f) encircle spins whose bond energies are affected by transition to a different ground state, and thus represent the domain walls.}
\end{figure}

To understand the very different behaviors of red and blue spin chains we consider the ground state degeneracies and the corresponding domain wall structures of these two weakly coupled spin systems (Fig.~\ref{fig:9}). For the red chains there are two ground states which are time-reversal partners. These are shown as $R1$ and $R1^*$ in Fig.~\ref{fig:9}(a). A transition between these structures involves a domain wall that costs a finite energy of $\Delta E=-2J_{r1}+4J_{r2}$. The situation is more complicated in the blue chains. In their ground state there are four sites per magnetic unit cell and therefore four different types of domains that correspond to shifting the ($\uparrow\uparrow\downarrow\downarrow$) motif with respect to the origin. We label these as $B1$, $B2$ and their time-reversal partners $B1^*$ and $B2^*$ in Fig.~\ref{fig:9}(d). Low energy domain walls correspond to transitions between ground states that are shifted by only one lattice site. Their ``chiral" character can be appreciated by comparing the energy cost of a domain wall that goes from $B1$ to $B2$ [counterclockwise in Fig.~\ref{fig:9}(e)] to a domain wall that effectuates a transition from $B2$ to $B1$ [clockwise in Fig.~\ref{fig:9}(e)]. By convention we choose the right direction as the positive direction of the spin chain. The domain walls are chiral because their energy depends on handedness in Fig.~\ref{fig:9}(e): $\Delta E_{\rm ccw}=J_{b1}-2J_{b2}$ and $\Delta E_{\rm cw}=-J_{b1}-2J_{b2}$. The larger ground state degeneracy and the correspondingly more complex domain wall structures for blue chains complicates their attainment of thermodynamic equilibrium.

It is useful to consult the 3D ANNNI model\cite{Selke1988213} to understand how the different domain wall structures may affect the low temperature magnetic ordering. For $J_{r2}/J_{r1}=-0.4(3)$, the mean field phase diagram of the 3D ANNNI model features a single phase transition from a paramagnetic (PM) phase to 3D N\'eel order, as we observe for the red chains in $\rm SrHo_2O_4$.  The exchange parameters of the blue chains [$J_{b2}/J_{b1}=1.5(3)$] however, place these in a complicated part of the 3D ANNNI phase diagram. Between the PM phase and 3D double-N\'eel order exists a large number of 3D LRO phases with different modulation wave vectors. These can be described in terms of different arrangements of domain wall defects within the double N\'{e}el structure. Effective interactions between defects stabilize these various phases at different temperatures.

With this picture in mind, the continuous peak center shift observed in Fig.~\ref{fig:8}(d) reflects domain wall rearrangement along blue chains. Focusing on domain walls, full 3D LRO requires registry in the placement of transitions between domains along all blue spin chains. Such collective domain wall motion requires rearrangement of large numbers of spins and so can be a slow process. Further, in the non-Kramers doublet ground state of Ho$^{3+}$, a spin flip can only take place through tunneling or a thermal process involving  higher CEF levels. In the recently proposed CEF level scheme\cite{Poole2014} blue sites have a large energy gap ($\sim$ 12 meV) to the first excited state (compared to $\sim$ 1 meV for red sites). This can be expected to reduce the tunneling and thermal rate for blue spin flips at low temperatures.

We now return to the important experimental observation that the spin configuration on blue chains ceases to evolve when red chains become fully ordered for $T<T_\mathrm{S}$ (Fig.~\ref{fig:8}). This indicates fluctuating exchange fields from spin dynamics in the red chains - which have a lower barrier to spin flips - are the dominant source of spin dynamics on blue chains. Ironically, it thus appears to be the development of saturated order on red chains that increases the relaxation time for domain wall motion in blue chains and stunts their inter-chain correlations.

In conclusion, our experiments on SrHo$_2$O$_4$ suggest a magnetically disordered state can persist in the low $T$ limit within a high quality crystal not because it is energetically favorable but because it is the thermodynamic equilibrium state when, upon cooling, ergodicity is lost.  $\rm SrHo_2O_4$ also illustrates the  remarkably disruptive impact of  topological defects in $d$ dimensions (here $d=1$) on $d+1$ dimensional order. Combining the ingredients of large scale emergent structures from frustration and reduced dimensionality, kinetically trapped spin disorder may actually be possible without quenched disorder.

\begin{acknowledgments}

We thank O. Tchernyshyov, J. Zang, Y. Wan for discussion. Work at IQM was supported by the US Department of Energy, office of Basic Energy Sciences, Division of Materials Sciences and Engineering under grant DE-FG02-08ER46544. This work utilized facilities supported in part by the US National Science Foundation under Agreement No. DMR-0944772. Research conducted at ORNL's High Flux Isotope Reactor was sponsored by the Scientific User Facilities Division, Office of Basic Energy Sciences, US Department of Energy. Ames Laboratory is operated for the U.S. Department of Energy by Iowa State University under Contract No. DE-AC02-07CH11358.
\end{acknowledgments}

\bibliography{Mrefs}

\end{document}